\documentclass[sigconf]{acmart}
\usepackage{graphicx}
\usepackage{balance} 
\usepackage{amsmath}
\DeclareMathOperator*{\argmax}{arg\,max}
\usepackage{nameref}
\usepackage{enumitem}
\usepackage{color, colortbl}
\usepackage{multirow, multicol}
\usepackage{subfigure}
\usepackage{caption}
\usepackage{algpseudocode}
\usepackage{amsfonts}
\usepackage{algorithm}
\usepackage{bm}
\usepackage{booktabs}
\usepackage{slashbox}
\usepackage{float}

\AtBeginDocument{%
  }

\setcopyright{acmlicensed}
\copyrightyear{2025}
\acmYear{2025}
\acmConference[EARL Workshop at RecSys'25]{ Workshop on Evaluating and Applying Recommendation Systems with
Large Language Models at RecSys ’25}{September 22--26, 2025}{Prague, Czech Republic}
\acmBooktitle{Proceedings of EARL
’25: Workshop on Evaluating and Applying Recommendation Systems with
Large Language Models at RecSys ’25}

\begin{document}

\title{SemSR: Semantics aware robust Session-based Recommendations}

\author{Jyoti Narwariya*}
\email{jyoti.narwariya@tcs.com}
\affiliation{ %
\institution{TCS Research}\city{New Delhi}\country{India}}
\author{Priyanka Gupta*}
\email{priyanka.g35@tcs.com}
\affiliation{%
\institution{TCS Research}\city{New Delhi}\country{India}}
\author{Muskan Gupta*}
\email{muskan.gupta4@tcs.com}
\affiliation{%
\institution{TCS Research}
\city{New Delhi}
\country{India}}
\author{Jyotsana Khatri}
\email{khatri.jyotsana@tcs.com}
\affiliation{%
\institution{TCS Research}\city{Pune}\country{India}}
\author{Lovekesh Vig}
\email{lovekesh.vig@tcs.com}
\affiliation{%
\institution{TCS Research}\city{New Delhi}\country{India}}

\renewcommand{\shortauthors}{Jyoti Narwariya*, Priyanka Gupta*, Muskan Gupta*, Jyotsana Khatri \& Lovekesh Vig}

\begin{abstract}
Session-based recommendation (SR) models aim to recommend items to anonymous users based on their behavior during the current session. While various SR models in the literature utilize item sequences to predict the next item, they often fail to leverage semantic information from item titles or descriptions impeding session intent identification and interpretability. Recent research has explored Large Language Models (LLMs) as promising approaches to enhance session-based recommendations, with both prompt-based and fine-tuning based methods being widely investigated. However, prompt-based methods struggle to identify optimal prompts that elicit correct reasoning and lack task-specific feedback at test time, resulting in suboptimal recommendations. Fine-tuning methods incorporate domain-specific knowledge but incur significant computational costs for implementation and maintenance.
In this paper, we present multiple approaches to utilize LLMs for session-based recommendation: (i) in-context LLMs as recommendation agents, (ii) LLM-generated representations for semantic initialization of deep learning SR models, and (iii) integration of LLMs with data-driven SR models. Through comprehensive experiments on two real-world publicly available datasets, we demonstrate that LLM-based methods excel at coarse-level retrieval (high recall values), while traditional data-driven techniques perform well at fine-grained ranking (high Mean Reciprocal Rank values). Furthermore, the integration of LLMs with data-driven SR models significantly outperforms both standalone LLM approaches and data-driven deep learning models, as well as baseline SR models, in terms of both Recall and MRR metrics.

\end{abstract}



\maketitle
\def\thefootnote{*}\footnotetext{These authors contributed equally to this work.}
\section{Introduction}
Session-based Recommendation (SR) models are becoming increasingly popular due to their ability to recommend items based only on interactions in the current session for anonymous users. Several SR models employing deep learning architectures are proposed in the literature \cite{hidasi2015session,wu2018session,gupta2021causer,gupta2019niser,hou2022core,kang2018self,li2017neural,liu2018stamp,xie2022contrastive}. 
Recently, motivated by the notable achievements of LLMs, several works try to employ these large models in recommendations \cite{hu2024enhancing, liu2024llm, wang2024re2llm, liu2025llmemb}. SAID \cite{hu2024enhancing} aims to learn item embeddings that align with the textual descriptions of items. Authors proposed a two-stage training scheme. At first stage, SAID employs a projector module to transform an item ID into an embedding and feeds it into an LLM to explicitly to elicit the item’s textual token sequence from the LLM. At second stage, embeddings are used for extracting the entire sequence’s representation for recommendation. LLM-ESR \cite{liu2024llm} obtains semantic embeddings of items and users by encoding prompt texts from LLMs. Authors devise a dual-view modeling framework that combines semantic and collaborative information. Specifically, the embeddings derived from LLMs are frozen to avoid deficiency of semantics. Next, they propose a retrieval augmented self distillation method to enhance the sequence encoder of an SR model using similar users. \cite{hu2025alphafuse} learns ID embeddings in the null space of language embeddings to combine semantic and collaborative knowledge in an optimal way.

In this work, our aim is also to utilize the capability of LLMs in SR models in an efficient and effective manner and propose \textbf{SemSR:} \textbf{Sem}antics aware robust \textbf{S}ession-based \textbf{R}ecommendations; a framework where LLMs capability is incorporated in the form of embedding of items generated using pre-existing LLM. Our approach is different from the existing above approaches as we train \textbf{SemSR} end to end unlike SAID which has two stages of learning, and LLM-ESR that use self distillation method to enhance the sequence encoder of an SR model.\\
Through extensive experiments on two real-world publicly available datasets, we demonstrate that 
LLM-based methods excel at coarse-level retrieval (high recall values), while traditional data-driven techniques perform well at fine-grained ranking (high Mean Reciprocal Rank values). Furthermore,
\textbf{SemSR} significantly outperforms both standalone LLM approaches and data-driven deep learning models, as well as baseline SR models, in terms of both Recall and MRR metrics.
 
Two distinct SR models are employed to show the efficacy of our approach i.e., MSGAT \cite{qiao2023bi}, and NISER \cite{gupta2019niser}.
However, the approach is model agnostic and can be employed for any SR model from the literature. 
\vspace*{-\baselineskip}

\section{Related Work}

\textbf{Session-based Recommendation:}
SR methods have evolved from traditional techniques such as Markov Chains \cite{jamali2010matrix, he2016fusing} and collaborative filtering \cite{ekstrand2011collaborative, wang2019collaborative}, to advanced deep learning-based approaches. Early methods struggled with capturing complex user behaviors and sequential dependencies, while deep learning models, such as GRU4Rec \cite{hidasi2015session}, SASRec\cite{kang2018self}, and SRGNN\cite{wu2018session}, have significantly improved predictive accuracy by leveraging sequential modelling, self-attention \cite{vaswani2017attention}, and graph-based representations \cite{wang2021graph,gupta2019niser,gupta2024scm4sr,gupta2024guided, qiao2023bi}. 

\textbf{LLM as Recommender Systems (RS):}
LLMs have recently demonstrated unparalleled capabilities in natural language understanding, reasoning and beyond \cite{zhao2023survey, min2023recent}. The idea of directly utilizing the LLMs as RS have gained significant traction with models such as GPT \cite{brown2020language}, BERT \cite{devlin2019bert}, and LLaMA \cite{touvron2023llama} which are trained on vast corpora of text, enabling a rich understanding of semantics. Due to their strong  understanding of language and context, LLMs can generate more personalized, and context-aware recommendations compared to conventional models. Recent works \cite{geng2022recommendation, dai2023uncovering,sanner2023large,ji2024genrec} have demonstrated that prompt-based LLMs can effectively recommend items by leveraging in-context learning. While prompt-based and fine-tuned LLM based techniques show strong potential, they are sensitive to the design of prompts and often lack up-to-date data knowledge which can lead to irrelevant recommendations in dynamic environments like e-commerce.

\textbf{LLMs to augment Recommender Systems:}

As more and more LLMs have been developed, the research has progressively explored how to utilize the knowlege in LLMs to improve the sequential recommendations. Recently, several works have highlighted the effectiveness of LLMs as components in recommendation tasks \cite{xi2024towards,he2023large,guo2024integrating,qiao2024llm4sbr,liu2024llm,hu2024enhancing,wang2024re2llm,liu2025llmemb}. KAR \cite{xi2024towards} proposes to use language embeddings as additional input in the learning of ID embeddings. LLM4SBR \cite{qiao2024llm4sbr} utilizes a a two step strategy. Firstly, session data is transformed into both textual and behavioral modalities, allowing LLMs to infer session intent from textual descriptions. Secondly, SR models use behavioral data to align and average session representations across two different modalities. LLM-ESR\cite{liu2024llm} retrieves semantic embeddings of items and users by encoding prompts from LLMs and uses a retrieval augmented self distillation method to enhance the sequence encoder of an SR model. SAID\cite{hu2024enhancing}, on the other hand, evolves a two-stage training process: the first stage involves generating item embeddings by leveraging the projector module and LLM, and in the second stage, learned item embeddings are input into the sequential model to extract the entire sequence’s representation for recommendation. In contrast to above methods, our proposed method SemSR introduces an end-to-end framework that directly incorporates LLM-generated item embeddings in SR model to obtain top-$K$ recommendations. AlphaFuse \cite{hu2025alphafuse} proposes an approach which injects collaborative signals into the null space of language embeddings which helps in preserving the semantic information. In AlphaFuse, trainable ID embeddings are learned in an orthogonal null space. Our approach is different, it tries to fuse semantic and collaborative signals (trainable) and the entire model is trained jointly.

\section{Proposed Framework} 

\begin{figure}[t]
\center
\vspace{-1em}
\subfigure[\centering SemSR-F]{\centering
\includegraphics[width=\linewidth,trim=1mm 5mm 2mm 2mm, clip=true]{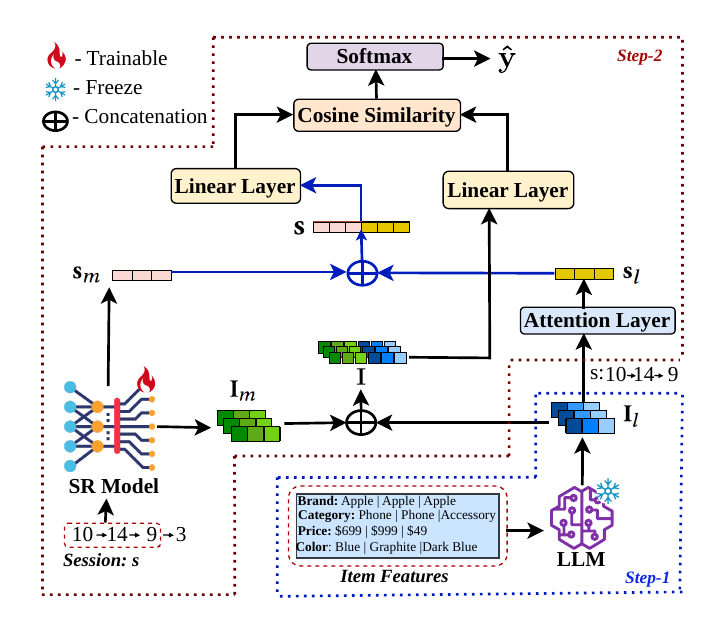}}
\vspace{-1em}
\subfigure[ SemSR-I]{\centering
\includegraphics[width=\linewidth,trim=0mm 18mm 17mm 2mm, clip=true]{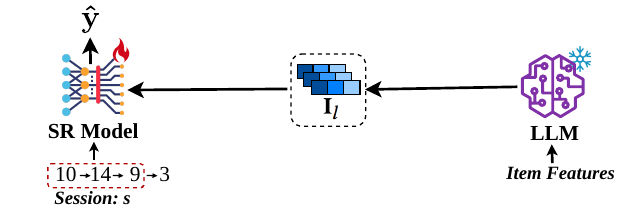}}

\caption{An overview of our proposed framework SemSR. (a) SemSR-F: For a given session s, SR model provides item ids based session representation $\textbf{s}_m$ and item representations $\mathbf{I}_m$. LLM model generates semantic item representations $\mathbf{I}_l$ for all items and $\textbf{s}_l$ is the LLM session representation obtained using embeddings of items $i_j$ where $\forall i_j \in s$ via attention layer. Further $\textbf{s}$ and $\mathbf{I}$, session and item representations obtained in end-to-end framework. Step-1 represents semantically aligned embedding learning and Step-2 represents agnostic sequential recommender training. (b) SemSR-I: Leverage LLM generated item representations $\mathbf{I}_l$ in SR model.\label{fig:semsr}} 
\vspace{-6mm}
\end{figure}

This section will introduce the problem definition and details of our approach:

\subsection{Problem Definition}
Suppose that $\mathcal{S}$ denotes the set of all sessions in the logged data containing user-item interactions (e.g. click/view/order), and $\mathcal{I}$ denotes the set of $n$ items observed in $\mathcal{S}$.
Any session $s \in \mathcal{S}$ is a sequence of item-click events: 
$s = (i_{s,1},i_{s,2},\ldots,i_{s,|s|})$, where  $i_{s,j}$ ($j=1\ldots |s|$) $\in$ $\mathcal{I}$, denotes the  $j^{th}$ clicked item in session $s$.
The goal of SR is to predict the next item $i_{s,|s|+1}$ as the target class in an $n$-way classification problem by estimating the $n$-dimensional item-probability vector $\mathbf{\hat{y}}_{s,|s|+1}$ corresponding to the relevance scores for the $n$ items. The $K$ items with the highest scores constitute the top-$K$ recommendations.

\vspace{-2mm}
\subsection{LLM as Recommender System}

\begin{figure}[!t]
    \centering
    \includegraphics[width=0.5\textwidth,height=52mm, trim=4mm 97mm 170mm 5mm, clip=true]{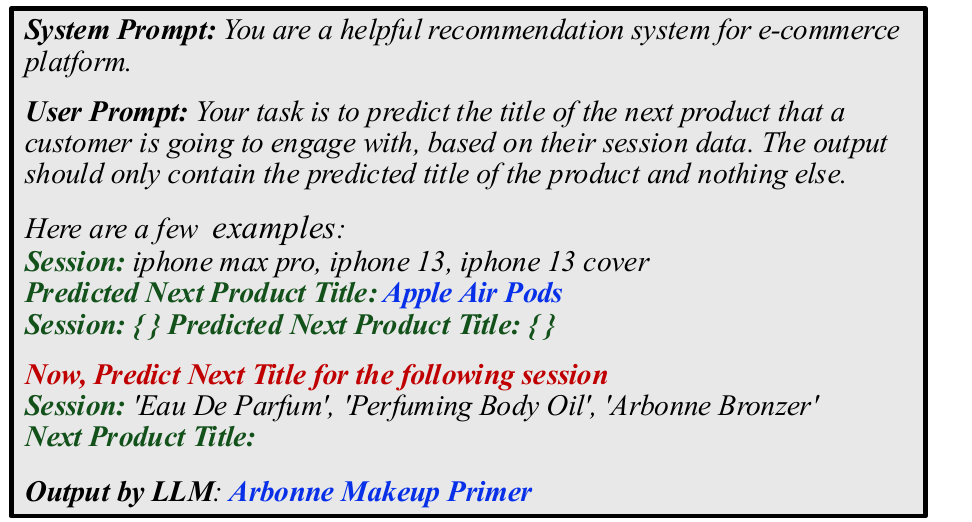}
    \caption{Few-shot in-context prompt (FS-LLM) template}
    \label{fig:prompt}
    \vspace{-7mm}
\end{figure}

We leverage an LLM model (Llama3-8B-Instruct\footnote{https://huggingface.co/meta-llama/Llama-3.1-8B-Instruct}) as an in-context text generation method to generate the title of the next recommended item for a given session which contains a sequence of clicked items titles. We use different in-context prompt design strategies such as few-shot in-context prompting, chain-of-thought (CoT) prompting to generate the next recommended title.

\textbf{Prompt Design:} We utilize different prompting strategies but use a common definition for designing prompts i.e., firstly, we define the role for system, then we define task for the model and get the response title from the LLM.
Different prompting strategies are as follows:

\begin{itemize}[leftmargin=*]
    \item \textit{Few-shot in-context prompt (FS-LLM):} In this prompt, in addition, we provide few shot input-output examples to the LLM, followed by the query session and allow the model to predict the next item title. The prompt template for FS-LLM is shown in figure \ref{fig:prompt}.
    \item \textit{Zero-shot Chain-of-Thought prompt (ZCoT-LLM):} In this prompt, we design a two step chain of thought prompt. First we generate rationale and then generate recommendations based on the generated rationale.
    \item \textit{Few-shot Chain-of-Thought prompt (FSCoT-LLM):} Similar to ZCoT-LLM strategy, here we also provide some examples in the prompt to learn input and output relations.
    
\end{itemize}

\textbf{LLM Inference:}
To retrieve top-k recommendations, we use  a vector database (Chromadb\footnote{https://docs.trychroma.com/docs/overview/getting-started}) to store item embeddings obtained from the Llama3 model. We then retrieve top-k recommendations based on cosine similarity between the generated next item title and stored items in the database. 
\vspace{-2mm}
\subsection{SemSR}

The overall architecture of SemSR is shown in Figure \ref{fig:semsr}. The following section specifies the implementation details of the architecture. 
\vspace{-3mm}
\subsubsection{Semantic-aware Item and Session Representation}
Each item is mapped to a $d_1$-dimensional vector from the trainable embedding look-up table to obtain interaction-based item embeddings as $\textbf{I}_m =[\textbf{i}_{m_{1}}, \textbf{i}_{m_{2}}, \dots, \textbf{i}_{m_{n}}]^T$ $\in$ $\mathbb{R}^{n \times d_1}$ 
such that each row $\textbf{i}_{m_j} \in \mathbb{R}^{d_1}$ is the $d_1$-dimensional embedding vector corresponding to item $i_j \in \mathcal{I}$ $(j=1, 2, ..., n)$. Further, each item is mapped to a $d_2$-dimensional vector by inputting item features like brand, category, price, color, title, description, etc to an LLM model (in this work, we consider Angle \cite{li2023angle} and LLM2Vec \cite{behnamghader2024llm2vec} embedding with BERT and LLama-3.1-8B as the backbone model respectively) and obtain LLM-based item embeddings as  $\textbf{I}_l=[\textbf{i}_{l_{1}}, \textbf{i}_{l_{2}}, ..., \textbf{i}_{l_{n}}]^T \in \mathbb{R}^{n \times d_2}$. 
These embeddings are frozen during model training.

We denote data-driven (interaction based) session embeddings by $\textbf{s}_m$ and LLM based session embedding by $\textbf{s}_l$. 
Consider any function $f$ (e.g., SR model's neural network from literature \cite{gupta2019niser,qiao2023bi}), parameterized by $\theta$ - maps sequence of items in session $s$ to $\textbf{s}_m = f(\textbf{I}_{m,s}, \theta)$, where, $\textbf{I}_{m,s} = [\textbf{i}_{m,s,1}, \textbf{i}_{m,s,2}, ..., \textbf{i}_{m,s,|s|}]^T \in \mathbb{R}^{|s| \times d_1}$. 
For obtaining $\textbf{s}_l$, we compute the soft-attention weight of the $j$-th item in session $s$ as $\alpha_j = \textbf{q}^T sigmoid(\textbf{W}_1\textbf{i}_{{l,s,|s|}} + \textbf{W}_2\textbf{i}_{l,s,j} + c)$, where $(j = 1, 2, ..., |s|-1)$, 
$\textbf{q}, \textbf{c} \in \mathbb{R}^{d_{2}}$, $\textbf{W}_1,\textbf{W}_2 \in \mathbb{R}^{d_2 \times d_2}$, and $\textbf{i}_{l,s,|s|}$ is the most recent item in session $s$.
The $\alpha_j$'s are further normalized using a softmax operation yielding intermediate session embedding $\textbf{s}' = \sum_{j=1}^{|s-1|}\alpha_j \textbf{i}_{l,s,j}$. The LLM-based session embedding $\textbf{s}_l$ is a linear transformation over the concatenation of intermediate session embedding $\textbf{s}'$ and the embedding of the most recent item $\textbf{i}_{l,s,|s|}$, s.t. $\textbf{s}_l = \textbf{W}_3[\textbf{s}';\textbf{i}_{l,s,|s|}]$, where $\textbf{W}_3 \in \mathbb{R}^{d_2 \times 2d_2}$. 

Finally, semantic aware session embeddings are obtained by linear transformation over the concatenation of $\textbf{s}_m$ and $\textbf{s}_l$ as $\textbf{s} = \textbf{W}_4[\textbf{s}_m;\textbf{s}_l$], where $\textbf{W}_4 \in \mathbb{R}^{d \times (d_1+d_2)}$. Further, semantic aware item embeddings are computed as linear transformation over the concatenation of $\textbf{I}_m$ and $\textbf{I}_l$ as $\textbf{I} = \textbf{W}_5[\textbf{I}_m;\textbf{I}_l$], where $\textbf{W}_5 \in \mathbb{R}^{d \times (d_1 + d_2)}$.
The semantic aware item and session embeddings are then used to obtain the relevance score for next clicked item $i_k$
computed as, 
\begin{equation}\label{eq:sc}
    \hat{\textbf{y}}_k = \frac{\exp{(\textbf{i}^T_k \textbf{s})}}{\sum^n_{j=1} \exp{(\textbf{i}^T_j \textbf{s})}}.
\end{equation}
\vspace{-2mm}
\subsubsection{Training and Inferencing of SemSR}
The goal is to obtain $\textbf{s}$ that is close to the embedding $\mathbf{i}_{s,|s|+1}$ of the target item $i_{k}=i_{s,|s|+1}$, where $k$ is estimated class for the target item,  $k = \argmax_j~ \mathbf{i}_j^T\mathbf{s}$ with $j=1\ldots n$. 
For this $n$-way classification task, softmax (cross-entropy) loss is used during training for estimating $\boldsymbol \theta$ by minimizing the sum of $\mathcal{L(\hat{\mathbf{y}})} = - \sum_{j=1}^{m}\mathbf{y}_j\text{log}(\hat{\mathbf{y}}_j)$
over all training samples, where $\mathbf{y} \in \{0,1\}^n$ is a 1-hot vector with  $\mathbf{y}_k=1$ corresponding to the correct (target) class $k$. 

During inference, the final recommendation scores for the $n$ items are computed by eq. \ref{eq:sc}. The top-$K$ items are considered as the recommended items.
\begin{table}[!ht]
\caption{Statistics of the datasets used for experiments.}
\label{tab:stats}
\begin{tabular}{lccccc}
\hline
{ \textbf{Datasets}} &{{ \textbf{\#train}}} & { \textbf{\#test}} & { \textbf{\#items}} & { \textbf{\begin{tabular}[c]{@{}c@{}}Avg. $|s|$\end{tabular}}} \\ \hline
{ Amazon-M2 (UK)}     & {{ 1172181}}          & { 115936}          & { 499611}           & { 4.12} \\ \hline
{ Amazon-Beauty}     & {{ 290512}}           & { 21580}           & { 54615}            & { 6.40}\\ \hline
\end{tabular}
\end{table}
\section{Experimental Evaluation}

\begin{table*}[!ht]

\caption{Evaluation on Amazon-M2(UK) and Beauty datasets. We report overall performance in terms of Recall@20, MRR@20, Recall@100 and MRR@100.  
Bold numbers are for the best.}
\label{tab:results}
\begin{tabular}{llcccccccc}
\hline
&                                   & \multicolumn{4}{c}{{\textbf{Amazon-M2 (UK)}}}                         & \multicolumn{4}{c}{{\textbf{Amazon-Beauty}}}                           \\  \cline{3-10} 
& \multirow{-2}{*}{\textbf{Method}} & {\textbf{R@20}} &{\textbf{MRR@20}}       & {\textbf{R@100}} & {\textbf{MRR@100}} & {\textbf{R@20}}  & {\textbf{MRR@20}} & {\textbf{R@100}} & {\textbf{MRR@100}} \\ \cline{1-10} & {NARM}                             & {32.49}                                  & {17.00}                                  & {42.55}                                  & {17.26}                                    & {15.68}                                  & {4.49}            & {29.36}          & {4.82}             \\ 
& {SRGNN}                             & {41.77}                                  & {23.54}                                  & {51.67}                                  & {23.54}                                    & {19.67}                                  & {6.14}                                    & {34.59}          & {6.05}             \\ \hline
                                & MSGAT       & 38.86          & 21.41          & 49.95          & 21.69          & 21.55          & 6.41          & 38.45          & 6.83          \\ \hline
\multirow{4}{*}{\begin{tabular}[c]{@{}l@{}}Angle embeddings \\ with BERT \end{tabular}} & SemMSGAT-I  & 46.00          & 14.90          & 63.49          & 15.33          & 15.30          & 3.93          & 32.06          & 4.32          \\ 
                                & SemMSGAT-I+ & 46.00          & 25.82          & 63.49          & 27.25          & 15.30          & 5.77          & 32.06          & 7.08          \\ 
                                & SemMSGAT-F  & 51.55          & 27.36          & 65.81          & \textbf{27.73} & 22.80          & 7.44          & 41.22          & 7.88          \\ 
                                & SemMSGAT-F+ & 51.55          & 27.37          & 65.81          & 27.48          & 22.80          & 7.66          & 41.22          & \textbf{7.94} \\ \hline
\multirow{4}{*}{\begin{tabular}[c]{@{}l@{}}Llama 3 \\embeddings\end{tabular}}& SemMSGAT-I  & 47.48          & 16.46          & 65.59          & 16.91          & 18.00          & 4.80          & 37.23          & 5.26          \\ 
                                & SemMSGAT-I+ & 47.48          & 26.15          & 65.59          & 27.45          & 18.00          & 6.19          & 37.23          & 7.32          \\ 
                                & SemMSGAT-F  & 52.66          & 26.12          & 68.82          & 26.53          & \textbf{24.07} & 6.93          & \textbf{43.64} & 7.40          \\ 
                                & SemMSGAT-F+ & 52.66          & 27.42          & 68.82          & 27.52          & \textbf{24.07} & \textbf{7.72} & \textbf{43.64} & { 7.85}    \\ \hline
                                & NISER       & 49.84          & 26.45          & 61.77          & 26.76          & 20.84          & 6.95          & 38.35          & 7.36          \\ \hline
\multirow{4}{*}{\begin{tabular}[c]{@{}l@{}}Angle embeddings\\  with BERT\end{tabular}} & SemNISER-I  & 46.96          & 14.25          & 64.44          & 14.68          & 19.62          & 5.07          & 37.00          & 5.49          \\ 
                                & SemNISER-I+ & 46.96          & 26.06          & 64.44          & 27.33          & 19.62          & 6.94          & 37.00          & 7.66          \\ 
                                & SemNISER-F  & \textbf{54.98} & 26.41          & \textbf{69.90} & 26.79          & 22.67          & 6.10          & 42.41          & 6.57          \\ 
                                & SemNISER-F+ & \textbf{54.98} & \textbf{27.71} & \textbf{69.90} & 27.57          & 22.67          & 7.33          & 42.41          & 7.63          \\ \hline
\multirow{4}{*}{\begin{tabular}[c]{@{}l@{}}Llama 3 \\embeddings\end{tabular}}& SemNISER-I  & 50.05          & 16.47          & 67.62          & 16.91          & 19.95          & 4.80          & 40.23          & 5.28          \\ 
                                & SemNISER-I+ & 50.05          & 26.85          & 67.62          & 27.57          & 19.95          &      6.54         & 40.23          &        7.38       \\ 
                                & SemNISER-F  & 54.26          & 25.19          & 69.87 & 25.59          & 20.95          & 5.42          & 40.36          & 5.89          \\ 
                                & SemNISER-F+ & 54.26          &  27.66              & 69.87 &       27.60         & 20.95          &   7.24           & 40.36          &  7.56            \\ \hline
\end{tabular}
\end{table*}

In this section, we conduct extensive experiments to answer the following research questions:
\begin{itemize}[leftmargin=*]
    \item \textbf{RQ1:} Are in-context LLMs' predictions competitive with data driven SR methods?
    \item \textbf{RQ2:} Does semantic-aware representations with data-driven models improve performance?
    \item \textbf{RQ3:} Does re-ranking top-K recommendations  improve performance in-terms of MRR? 

\end{itemize}

We consider two SR models from the literature to demonstrate the efficacy of \textbf{SemSR}, i.e., MSGAT \cite{qiao2023bi}, and NISER \cite{gupta2019niser}. 
We compared \textbf{SemSR} with exiting SR models from literature e.g., NARM \cite{li2017neural} and SRGNN \cite{belieni2025srgnn}.\\

\textbf{Dataset Details:}
We consider two datasets i.e., english dataset from Amazon KDDCup challenge 2023 (AmazonKDD-M2 (UK))\footnote{https://www.aicrowd.com/challenges/amazon-kdd-cup-23-multilingual-recommendation-challenge} and Amazon Beauty 2014 review dataset (Beauty) \footnote{https://cseweb.ucsd.edu/~jmcauley/datasets/amazon/links.html} to evaluate our approach. After preprocessing, the statistics of the datasets are shown in table \ref{tab:stats}. 

\textbf{AmazonKDD-M2 (UK):} 
We use a real-world multilingual recommendation dataset from KDDCup challenge 2023 to evaluate the effectiveness of SemSR. The dataset provides items and sessions in six different languages but for this paper non-english sessions are filtered out yielding $4,99,611$ items with average sessions length of $4.12$.
We consider data from task 1 of Amazon KDDCup challenge 2023. The number of given training/testing sessions considered are $11,72,181$/$1,15,936$ respectively. 
For LLM as RS experiments, we selected a subset of the data to minimize cost. We selected the most recent $10k$ sessions as the test set based on chronological splits to minimize the computational cost. 

\textbf{Amazon-Beauty:} We consider the Amazon-Beauty sub-category from the Amazon 2014 review dataset which 
contains timestamped user-item interactions from May 1996 to July 2014, and metadata contains items’ title, descriptions, categories, brands, price, etc. 
We only consider users having more than 5 reviews, filtering out less popular items that have a frequency less than 5 and removed sessions of length less than 2 from the data. We split sessions based on user ids and consider $80\%$:$10\%$:$10\%$ split of data into training, validation and test set, respectively. Average sessions length is $6.40$. Further, we create incremental sessions to improve training of the SR model.\\
\textbf{SemSR and its variants:}
We proposed the following variants of our approach:
\begin{itemize}[leftmargin=*]
    \item \textit{Semantic Initialization via LLM embeddings} (SemSR-I): Item embeddings are initialized by LLM embeddings for MSGAT \cite{qiao2023bi}, and NISER \cite{gupta2019niser} 
    (denoted as SemMSGAT-I, SemNISER-I) 
    respectively.
    \item \textit{Semantically initialized model fused with SR} (SemSR-F): LLM based item embeddings  are concatenated with data driven item embeddings followed by a linear transformation. Similarly,  LLM based session embedding are fused with data driven session embeddings. SemMSGAT-F, SemNISER-F 
    are variants for MSGAT, and NISER, 
    respectively.
    \item SemSR-I+, SemSR-F+: Re-ranked recommendations list obtained from SemSR-I, and SemSR-F using NISER model. 
\end{itemize}

\begin{table*}[!t]
\caption{LLM as RS comparison with data-driven SR methods on subset of Amazon-M2(UK) and Beauty dataset. \label{tab:llm}}
\begin{tabular}{lcccccccc}
\hline
\multicolumn{1}{c}{}                         & \multicolumn{4}{c}{\textbf{Amazon-M2 (UK)}}    & \multicolumn{4}{c}{\textbf{Amazon-Beauty}}              \\ \cline{2-9} 
\multicolumn{1}{c}{\multirow{-2}{*}{\textbf{Method}}} & \textbf{R@20}  & { { \textbf{MRR@20}}} & { { \textbf{R@100}}} & { {\textbf{MRR@100}}} & \textbf{R@20}           & { { \textbf{MRR@20}}} & { { \textbf{R@100}}} & { {\textbf{MRR@100}}} \\ \hline
FS-LLM& 31.83 & 9.92 & 47.50& 10.30 & 7.07           & 1.90 & 14.48& 2.07    \\ 
ZCoT-LLM                                     & 25.16 & 7.65 & 39.29                              & 7.99  & 7.12           & 1.82 & 15.08& 1.99   \\ 
FSCoT-LLM                                    & 31.70 & 10.00& 47.97                              & 10.44 & 7.04           & 1.86 & 14.32& 1.89   \\ \hline
MSGAT                                        & 26.28 & 14.90                               & 34.74                              & 15.09                                & 21.55          & 6.41                                & 38.46& 6.83  \\ 
SemMSGAT-I                                   & 44.28                         & 14.76& 61.09& 15.18 & 18.00          & 4.80                                & 37.23                              & 5.26                                 \\ 
SemMSGAT-F                                   & \textbf{47.64}                & \textbf{21.55}& 63.79& \textbf{21.96}& \textbf{24.07} & 6.93                                & \textbf{43.64}                     & \textit{\textbf{7.40}}               \\ \hline
NISER                                        & 38.02                         & 20.21& 47.56& 20.46 & 20.84          & \textbf{6.95}                       & 38.35                              & 7.36                                 \\ 
SemNISER-I                                   & 46.64                         & 14.77& 62.97& 15.18 & 19.95          & 4.80                                & 40.23                              & 5.28                                 \\
SemNISER-F                                   & 47.17                         & 20.83& \textbf{63.83} & 21.25 & 20.95          & 5.42                                & 40.36                              & 5.89                                 \\ \hline
\end{tabular}
\end{table*}
\textbf{Evaluation Metrics:}
We use the standard offline evaluation metrics Recall$@K$ and Mean Reciprocal Rank (MRR$@K$). 
\textbf{Recall$@$K} represents the proportion of test instances which has the desired item in the top-K items. \textbf{MRR$@$K} (Mean Reciprocal Rank) is the average of reciprocal rank of desired item in recommendation list.\\ 
\textbf{Hyperparameter Setup:}
We use validation data for hyperparameter selection using Recall$@100$ as the performance metric for all approaches except in-context LLMs based approach as this does not require model training. We use the adam optimizer with mini-batch size $100$, momentum $0.9$, $d_1 = 100$, $d_2 = 1024$, and $d=100$. For NISER and its SemSR variants, we use the same parameters \cite{gupta2019niser} i.e., scaling factor $= 16.0$ and learning rate $0.001$. For MSGAT, 
we grid-search over learning rate $in$ \{$(0.1, 0.01, 0.001, 0.0001)$\}. The best learning rate on the validation set is $lr=0.001$. 

\subsection{Results and Discussion}
\begin{figure}
    \centering
    \includegraphics[width=0.32\textwidth]{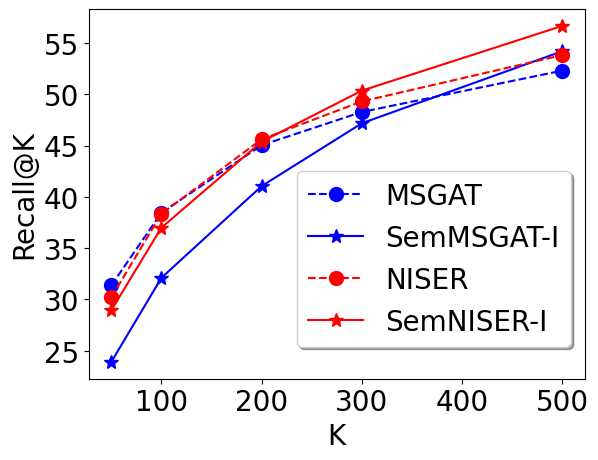}
    \vspace{-3mm}
    \caption{Recall@K with varying K indicating larger gains by using SemSR-I  over SR with increasing K on Amazon-Beauty dataset.}
    \label{fig:my_label}
    \vspace{-3mm}
\end{figure}
From \ref{tab:results}, we observe that MSGAT, and NISER 
performs significantly better that existing baselines NARM and SRGNN.

\begin{itemize}[leftmargin=*]
    \item \textbf{RQ1: LLM as RS vs data-driven SR model,}
    In table \ref{tab:llm}, we compare LLM as recommendation methods FS-LLM, ZCOT-LLM, and FSCoT-LLM with traditional data-driven SR models. We observe that while LLM as RS methods demonstrate some capability in recommendation task, their performance significantly lags behind data-driven models such as MSGAT and NISER. 
    For example, FSCoT-LLM achieves R@20 and MRR@20 score of 31.70 and 10.00, respectively on the Amazon-M2 dataset, and only 7.04 and 1.86, respectively on Amazon-Beauty dataset. These results suggest that although LLMs may offer generalization benefits, they currently lack the collaborative knowledge which is necessary for session-based recommendation.
    
    \item \textbf{RQ2: Effectiveness of Semantic Integration,}
    From table \ref{tab:results}, we observe that semantic initializations of embeddings significantly enhances performance at a coarse level of retrieval, i.e., 
    SemMSGAT-I and SemNISER-I 
    perform significantly better than respective vanilla SR models in terms of recall on the Amazon-M2 (UK) dataset and are comparable on the Amazon-Beauty dataset. 
    From table \ref{tab:results}, we also observe that fusion-based models SemMSGAT-F and SemNISER-F consistently outperform their vanilla SR models 
    as well as other existing methods in terms of recall as well as MRR which further emphasizes the advantages of incorporating semantic information into the data-driven SR models.  
    \item \textbf{RQ3: Assessing the benefits of Re-ranking,} 
    From table \ref{tab:results}, we observe that re-ranking using vanilla SR methods helps to improve fine-grained ranking i.e., it further improves MRR for all semantic variants SemMSGAT-I+, SemMSGAT-F+, SemNiser-I+ and SemNiser-F+ over SemMSGAT-I, SemMSGAT-F, SemNiser-I and SemNiser-F, respectively.
    \item From figure {\ref{fig:my_label}}, we observe that for lower $K$ value (50 and 100), vanilla SR models MSGAT and NISER perform better than MSGAT-I and NISER-I, respectively. However, for higher $K$ values (K=200, ..., 500) trend is reversed. Moreover, the performance gap between MSGAT vs MSGAT-I and NISER vs NISER-I widens as K increases. This suggests that semantic initialization via LLM embeddings excel at coarse-level retrieval.
\end{itemize}

\section{Conclusion}
In this work, we highlighted various methods of leveraging LLMs for SR, i.e., in-context prompting and retrieval via LLMs, and integrating LLM-based embeddings with deep learning based state-of-the-art SR models i.e., MSGAT \cite{qiao2023bi} and  NISER \cite{gupta2019niser}. 
We showed the comparison of different approaches incorporating LLMs, and demonstrated that LLMs can help to improve the performance of SR models by understanding semantics of items and their features/meta-information. 

\bibliographystyle{ACM-Reference-Format}
\balance
\bibliography{sigir}

\end{document}